\documentclass{vgtc}                          

\ifpdf
  \pdfoutput=1\relax                   
  \usepackage{graphicx}                
  \DeclareGraphicsExtensions{.pdf,.png,.jpg,.jpeg} 
\else
  \usepackage{graphicx}                
  \DeclareGraphicsExtensions{.eps}     
\fi%

\graphicspath{{figures/}{pictures/}{images/}{./}} 

\usepackage{wrapfig}
\usepackage{enumitem}
\usepackage{soul}
\usepackage{subfigure}
\usepackage{graphbox}
\usepackage[T1]{fontenc}
\usepackage[utf8]{inputenc}
\usepackage{flushend}
\usepackage{amsmath}
\usepackage{amssymb}

\usepackage{float}

\usepackage{microtype}                 
\PassOptionsToPackage{warn}{textcomp}  
\usepackage{textcomp}                  
\usepackage{mathptmx}                  
\usepackage{times}                     
\usepackage{cite}                      
\usepackage{tabu}                      
\usepackage{booktabs}                  

\usepackage{amsmath}
\newcommand{\setmeterb}[2]{\ensuremath{%
  \vcenter{\offinterlineskip
            \vspace{-4pt}
    \halign{\hfil##\hfil\cr
            $\scriptstyle#1$\cr
            \noalign{\vskip1pt}
            $\scriptstyle#2$\cr}\vspace{-2pt}
  }
  }%
}

\usepackage{todonotes}
\usepackage{pdfpages}
\usepackage{chngcntr}
\usepackage{makecell}

\usepackage{calc}
\newlength\myheight
\newlength\mydepth
\settototalheight\myheight{Xygp}
\PassOptionsToPackage{svgnames, x11names, dvipsnames}{xcolor}
\definecolor{whole-note}{HTML}{6A1B9A}
\definecolor{half-note}{HTML}{a33a95}
\definecolor{quarter-note}{HTML}{c96196}
\definecolor{eighth-note}{HTML}{e58d9b}
\definecolor{16th-note}{HTML}{f7bba5}
\definecolor{32nd-note}{HTML}{ffecb3}
\definecolor{half-rest}{HTML}{39A247}

\usepackage[most]{tcolorbox}
\newtcbox{\inlinebox}[1][]{box align=bottom,
 nobeforeafter,
 colback=#1,
 size=small,
 grow to left by=-1pt,
 grow to right by=-1pt,
 boxrule=0pt,
 sharp corners}
 \newtcbox{\inlineboxwithtext}[1][]{tcbox raise=-5pt,
 nobeforeafter,
 colback=#1!20,
 colframe=#1!85!black,
 size=fbox,
 boxrule=1pt,
 enlarge by=1pt,
 enlarge top by=0pt,
 sharp corners}

\newcommand{\subhead}[1]{\vspace{2pt} \noindent \textbf{#1}}

\newcommand{\noviceA}{$N_1$} 
\newcommand{\noviceB}{$N_2$} 
\newcommand{\noviceC}{$N_3$} 
\newcommand{\noviceD}{$N_4$} 

\newcommand{\expertA}{$E_1$} 
\newcommand{\expertB}{$E_2$} 
\newcommand{\expertC}{$E_3$} 
\newcommand{\expertD}{$E_4$} 

\newcommand{\MSA}{$MS_1$}
\newcommand{\MSB}{$MS_2$}

\onlineid{0}

\vgtccategory{Research}

\vgtcinsertpkg

\title{Augmenting Sheet Music with Rhythmic Fingerprints}

\newcommand\acc[1]{\hspace{0.02cm}\hspace{-0.04cm}#1}

 \author{
 Daniel Fürst\thanks{e-mail: firstname.lastname@uni-konstanz.de}\\ %
       \parbox{1.5in}{\scriptsize \centering University of Konstanz }%
\and 
Matthias Miller\acc{$^*$}\\ %
       \parbox{1.5in}{\scriptsize \centering University of Konstanz }%
\and 
Daniel A. Keim\acc{$^*$}\\ %
       \parbox{1.5in}{\scriptsize \centering University of Konstanz }%
\and 
Alexandra Bonnici\thanks{e-mail: alexandra.bonnici@um.edu.mt}\\ 
     \parbox{1.5in}{\scriptsize \centering University of Malta }%
\and 
Hanna Schäfer\acc{$^*$}\\ %
       \parbox{1.5in}{\scriptsize \centering University of Konstanz }%
\and 
Mennatallah El-Assady\acc{$^*$}\\ %
       \parbox{1.5in}{\scriptsize \centering University of Konstanz }
 }

\teaser{
  \centering
  \includegraphics[width=\linewidth]{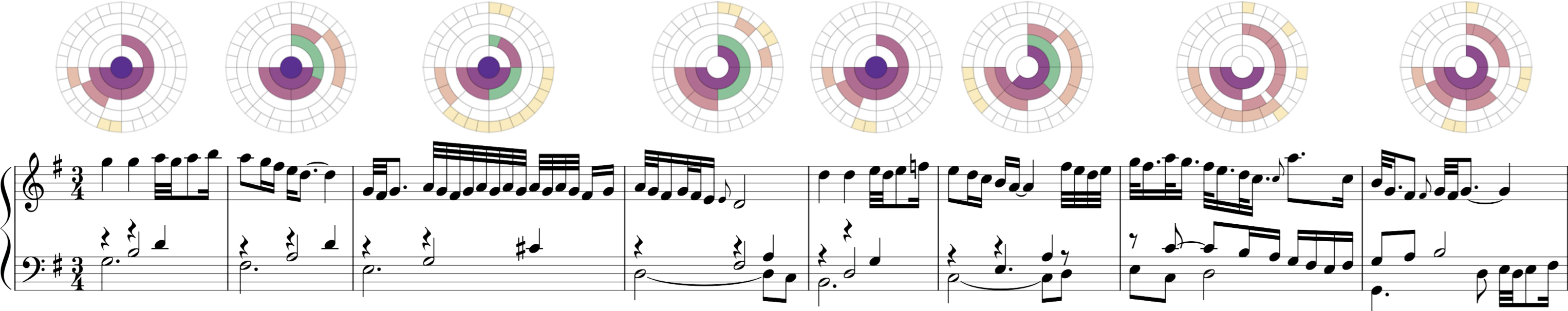}
  \caption{
Augmenting sheet music with rhythmic fingerprints allows for viewing only the rhythmic aspects of a composition.
  This excerpt shows the beginning of the \textit{Aria} from Bach's \textit{Goldberg Variations} conveying a similar rhythm in the first and fifth measure. 
  }
  \label{fig:teaser}
}

\abstract{
	In this paper, we bridge the gap between visualization and musicology by focusing on rhythm analysis tasks, which are tedious due to the complex visual encoding of the well-established Common Music Notation (CMN). Instead of replacing the CMN, we augment sheet music with rhythmic fingerprints to mitigate the complexity originating from the simultaneous encoding of musical features. The proposed visual design exploits music theory concepts such as the rhythm tree to facilitate the understanding of rhythmic information. Juxtaposing sheet music and the rhythmic fingerprints maintains the connection to the familiar representation. To investigate the usefulness of the rhythmic fingerprint design for identifying and comparing rhythmic patterns, we conducted a controlled user study with four experts and four novices. The results show that the rhythmic fingerprints enable novice users to recognize rhythmic patterns that only experts can identify using non-augmented sheet music.
} 

\makeatletter
\edef\orig@output{\the\output}
\output{\setbox\@cclv\vbox{\unvbox\@cclv\vspace{0pt plus 20pt}}\orig@output}
\makeatother

\begin{document}

\firstsection{Introduction}

\maketitle

Common Music Notation (CMN) resulted from a century-long development~\cite{Strayer2013} to visually encode musical information. 
While alternative notations have been proposed~\cite{Miller2018}, none has been adopted by the community to replace the CMN.
Hence, musicians, composers, and music analysts must be proficient at reading CMN~\cite{hultberg_approaches_2002}.
Inexperienced music readers face a steep learning curve to read sheet music.
Moreover, music analysis tasks require knowledge beyond the reading of individual notes.
Such tasks involve finding harmonic progressions~\cite{malandrino_visualization_2018}, melodic motifs, and rhythmic patterns~\cite{Toussaint2020} which helps understanding musical structure and interpreting a composition.
While melody, harmony, and rhythm are simultaneously present in music, analysts often need to focus on the different features separately~\cite{Miller2018}. 
In this paper, we concentrate on \textit{rhythm},
a compound feature building on the primitive attributes meter, onset, and duration~\cite{Stone1963}. 
Rhythmic characteristics build the foundation of compositions and have a significant influence on their organizational structure~\cite{Toussaint2020}. 
Besides, rhythm plays a crucial role in different music analysis tasks, including comparative and structure analysis.

Information Visualization has proven effective for music analysis tasks, such as identifying similar patterns or highlighting relevant aspects~\cite{Khulusi2020}.
In previous work, we demonstrate how augmenting sheet music with \textit{harmonic} fingerprints helps to identify patterns~\cite{Miller2019}. 
The results of the accompanying study confirm that with the harmonic fingerprints, even novice users could uncover harmonic patterns that only music experts were able to see before.
As both harmony and rhythm are essential for music analysis,
we argue that rhythm deserves separate consideration. 
Hence, as a direct follow-up, we introduce a rhythmic fingerprint to extend our work on augmenting sheet music.

In this paper, we contribute a \textit{rhythmic fingerprint} design to augment sheet music (see~\autoref{fig:teaser}) with entities that visually represent rhythmic characteristics. 
We exploit the visual metaphor of a clock using a radial tree layout to reflect the hierarchical structure of note durations. 
Through the augmentation,
we provide music readers with additional information that supports the identification of rhythmic relations in a composition. 
To evaluate the introduced approach and the usability of the rhythmic fingerprint, we conducted a user study with both novices and music experts. 
\section{Related Work} \label{sec:related-work}
Rhythm is one of the essential features of music~\cite{Krumhansl2000, Gouyon2005, Thul2008}. 
Understanding its characteristics and relations is crucial for music analysis tasks 
such as structure analysis, pattern identification, and interpretation~\cite{Swanwick1994}. 
During such analytical processes, the structure of music can reveal rhythmic relations. 
However, analyzing rhythm and its reflection in the music's structure requires to be proficient in music theory, 
which is a prerequisite that challenges novices and even intermediately-experienced musicians~\cite{Chan2007}. 
Rhythm visualizations offer one way to overcome this obstacle, exploiting the human's visual cognition ability. 
The augmentation of CMN with abstract visualizations enables close and distant reading~\cite{Janicke2015}.
In previous work, we applied such a combination to aid harmony analysis with \textit{harmonic fingerprints}~\cite{Miller2019} visually.

\subhead{Music Structure Analysis} -- 
A common \textit{music structure analysis} task is to extract the temporal sections, often referred to by lettered labels (i.e., $A, B, C, ...$)~\cite{Mueller2015MusicStructure}. 
These sections correspond to parts of the composition such as introduction, exposition, or coda and build the musical form. 
The so-derived structure segmentation of music depends on musical features such as harmony, rhythm, and melody~\cite{Mueller2015MusicStructure}. 
While key changes and modulations provide an obvious partitioning~\cite{dannenberg2008music}, towards phrase endings, particularly of major sections, music tends to slow down, i.e., note durations become longer. Also, musical phrases often start with similar rhythmic patterns that indicate suitable segmentation split points.
Consequently, the rhythmic qualities of a piece of music merit particular attention.
Hence, it is often possible to utilize rhythm to infer the structure of a musical piece~\cite{Leve2011}.
Boundaries that constitute the temporal segments in the music structure frequently correlate with rhythmic changes as in the occurrence of novel rhythmic patterns~\cite{Jensen2005,Mueller2015MusicStructure}. 
On the other hand, the repetition of rhythmic patterns helps to identify the musical contents that belong to one section~\cite{Mueller2015MusicStructure}. 
Radial hierarchy visualizations are particularly suitable to aid analysts in locating novelty and repetitions due to the hierarchy exhibited by rhythm and its cyclical appearance~\cite{Longuet-Higgins1979,Sethares2007,Draper2009}.

\subhead{Radial Hierarchy Visualization} -- 
To visualize the hierarchical relationship between different durations occurring in rhythmic patterns, Longuet-Higgins coined the \textit{rhythm tree}~\cite{Longuet-Higgins1979}. 
The hierarchical data structure within such trees is often visualized through \textit{tree maps}~\cite{Johnson1991} or \textit{Icicle plots}~\cite{Kruskal1983}. 
While these approaches effectively utilize space through their squared appearance~\cite{Johnson1991,DeWetering2020}, they are not suitable to represent the temporal aspect of rhythm accurately.
To classify joint visualizations of hierarchical and time-series data, Draper et al. devise a taxonomy for radial hierarchy visualizations~\cite{Draper2009}. 
Before distinguishing them according to their exhibited design pattern, visualizations are divided into three categories: \textit{Polar Plot}, \textit{Ring}, and \textit{Space Filling}. 
The \textit{Polar Plot} radiates lines that convey semantics from an origin to display the relationships between these branches. 
In contrast, \textit{Ring} visualizations display nodes on the circumference of a circle and highlight their relationships through connecting lines.

For the design proposed in this paper, we employ a space-filling visualization to convey the semantics denoted to the hierarchy and to enable comparison.
Schulz et al. concisely describe these requirements to generate a tree-layout identical to a \textit{fan chart}~\cite{Draper2009}~\cite{Schulz2013}. 
Draper et al. assign the visualization to their third category, \textit{Space-Filling}, within the \textit{concentric pattern}~\cite{Draper2009} similar to \textit{Sunburst} visualizations~\cite{Stasko2000}.
We transfer the concept of fan charts to the duration hierarchy since their structure is almost identical. 

\subhead{Rhythm Visualization} -- 
It is possible to visualize and analyze rhythmic aspects from musical compositions on several levels of abstraction~\cite{Janicke2015}. 
While aggregated visualization techniques facilitate retrieving an overview of the underlying data, they struggle with conveying the details.
In Digital Humanities, scholars often carefully examine data to understand its meaning and underlying relations on the close and distant reading level~\cite{Cheema2016}.

Typically, readers of music perform close reading using the CMN that encodes all rhythmic details~\cite{Miller2018}.
For instance, the \textit{binary string} as a rudimentary visual representation can aid the analysis of rhythm by considering the presence or absence of sound~\cite{Liu2012}.
This idea has been improved by the \textit{Time Unit Box System}~\cite{Toussaint2005} and its generalization, \textit{Drum Tablature}~\cite{Sethares2007}.
These alternative representations allow music analysts to focus on rhythmic characteristics at a close reading level.
Detaching such alternative representations hinders users in making a connection to the well-established CMN. 
Another approach by Robledo exploits texture in the background of the CMN, which is highlighting rhythmic units~\cite{robledo2010method} to improve their separation.
These concepts use a linear sequence to communicate the rhythmic aspects visually but do not support the cyclic nature of rhythmic patterns~\cite{Sethares2007} or their symmetry~\cite{Liu2012}.
Circular representations are more suitable to address these two characteristics~\cite{Sethares2007, Liu2012}. 

For instance, the \textit{necklace} notation displays rhythmic events on the circumference of a circle~\cite{Sethares2007}.
By exploiting the metaphor of a clock~\cite{Guo2017}, it supports the intuition of the connection between rhythmic repetitions and time.
This way of presentation is suitable for unveiling the regularity inherent to many rhythmic patterns~\cite{Sethares2007}. 
Similarly, the \textit{Polygon} notation represents beats in music by connected lines. 
These connections build up polygons enabling a comparison of rhythmic patterns. 
Such abstract visualizations facilitate the understanding of underlying patterns but aggravate the connection to the original representation. 
In the end, both close and distant reading is essential when it comes to reading music. 
Consequently, it is desirable to find solutions that support both concepts without sacrificing each other's advantages.

\subhead{Augmenting Sheet Music} -- 
In a review of visualizations for textual data, Jänicke et al. emphasize the importance of combined close and distant reading~\cite{Janicke2015}. 
While the latter cannot replace the former, distant reading can point the reader towards interesting spots, e.g., patterns in the data. 
Regarding the discussed abstract visualizations, there is a lack of such a connection between them and the CMN. 
Musicians usually achieve such a combination by augmenting visualizations above or below the sheet music systems, for example, through syllables~\cite{Dalby2005, Ester2006}. 
In the \textit{Generative Theory of Tonal Music} (GTTM), Lerdahl et al. provide numerous annotations through geometrical shapes, brackets, and trees to visualize rhythm features, such as articulation and meter~\cite{Lerdahl1996}. 
While the elaborated approaches provide an abstraction over the CMN, they are hard to compare due to their lack of conventional structure and color, using only black and white~\cite{Lerdahl1996, Dalby2005, Ester2006}. 
Besides, their abstraction is either too narrow (e.g., syllables) or too general (e.g., multi-measure tree-like structures in GTTM). 
To cope with this challenge, we propose a compromise by introducing an abstract fingerprint visualization for the musical feature of rhythm, which can be attached to sheet music to ensure the visual connection to the CMN.
\newcommand{\arrowspace}{\hspace*{-.03in}}
\newcommand{\arrowheight}{0.14in}
\newcommand{\graphheight}{1.45in}

\begin{figure*}[t]
\begin{minipage}{\textwidth}
    \subfigure[][
        The \textit{rhythm tree} \cite{Longuet-Higgins1979} puts the duration of notes and rests within measures into relation.
    ]{
    \includegraphics[align=c,height=\graphheight]{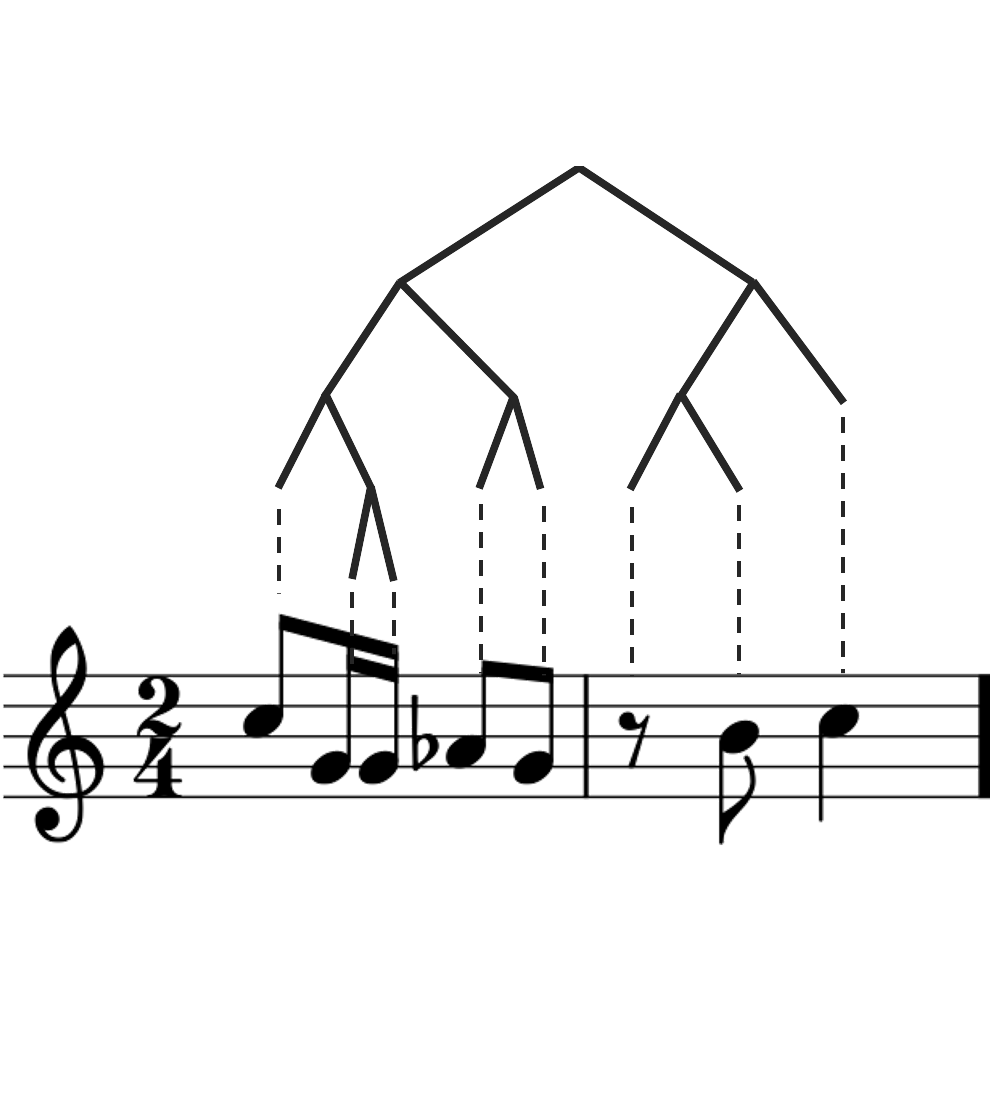}
    \label{fig:rhythm-tree-original}}
  \arrowspace
  \includegraphics[align=c,height=\arrowheight]{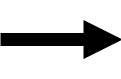}
  \arrowspace
    \subfigure[][
    This rhythm tree shows the relationship between subsequent duration layers for \textit{simple time signatures} only.
    We use two \textit{complementary} color scales 
    that preserve the duration order for rests and notes. 
            ]{
    \includegraphics[align=c,height=\graphheight]{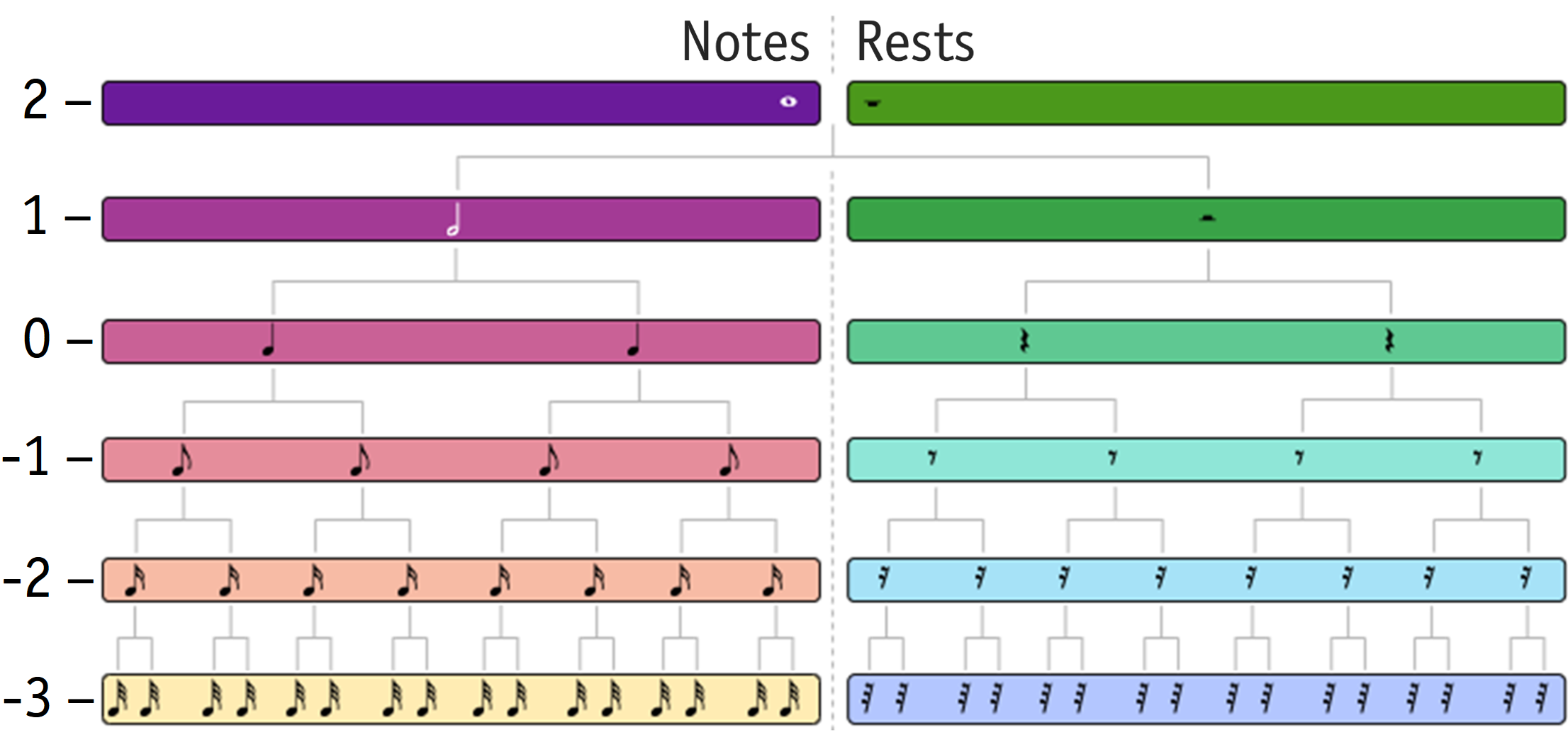}
            \label{fig:creating-rhythmic-fingerprint-rhythm-tree}}
  \arrowspace
  \includegraphics[align=c,height=\arrowheight]{img/arrow.png}
  \arrowspace
    \subfigure[][
    A \textit{skeleton} of the rhythmic fingerprint.
    ]{
    \includegraphics[align=c,height=\graphheight]{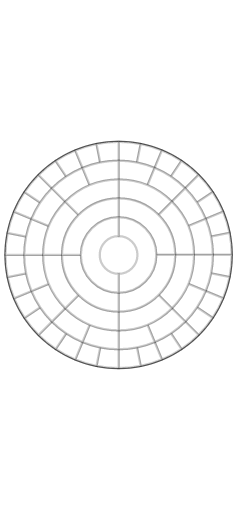} \label{fig:creating-rhythmic-fingerprint-empty}}
  \arrowspace
  \includegraphics[align=c,height=\arrowheight]{img/arrow.png}
  \arrowspace
      \subfigure[][
    \textit{Three voices} in a measure. 
    Notes and rests are colored according to (b).
    ]{\includegraphics[align=c,height=\graphheight]{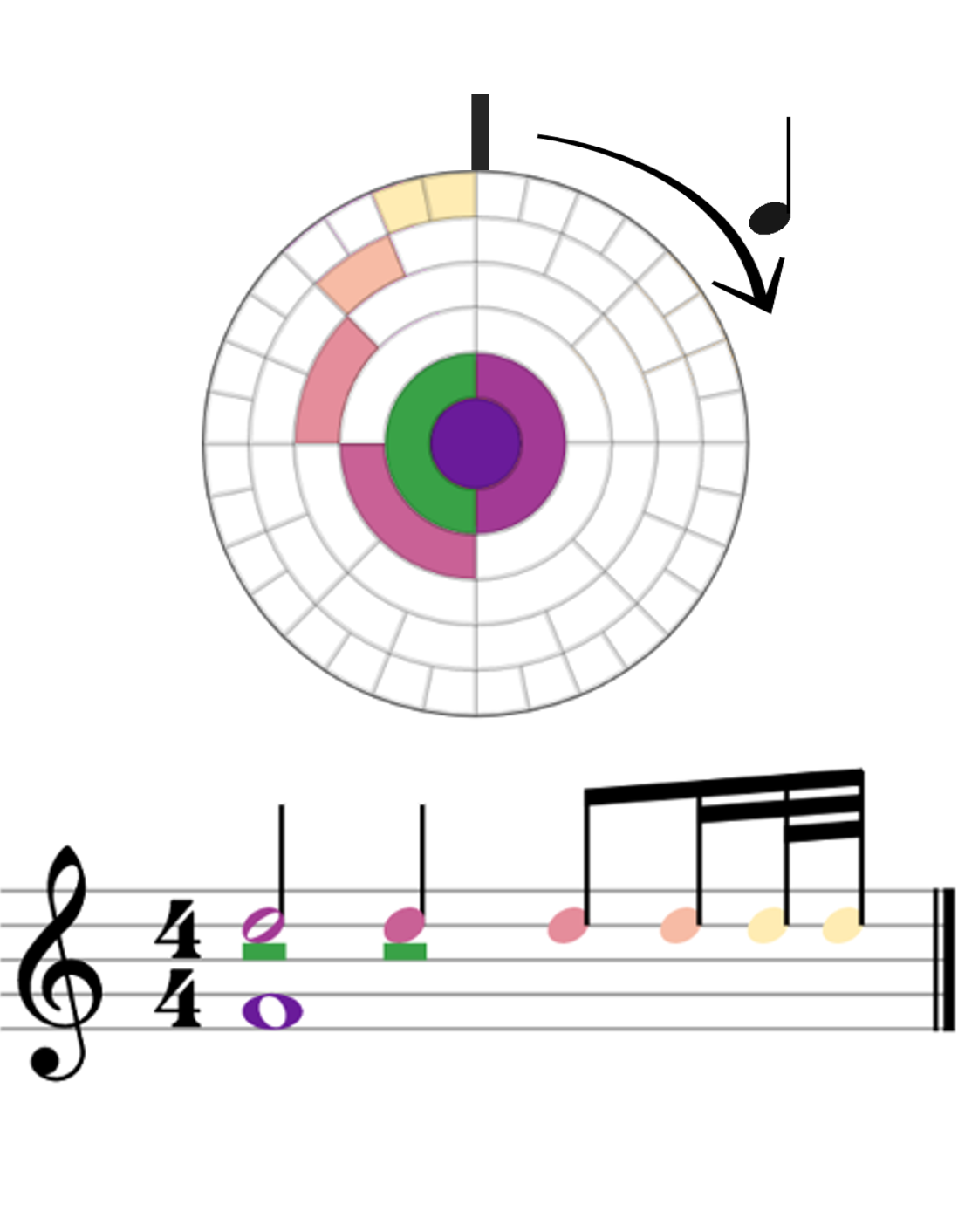}
    \label{fig:creating-rhythmic-fingerprint-measure}}
\end{minipage}
    \label{fig:creating-rhythmic-fingerprint}
    \caption{
    We utilize (a) the hierarchy present in the rhythm tree together with (c) the clock metaphor~\cite{Guo2017} to create a space-filling concentric visualization, the \textit{rhythmic fingerprint}. 
    (d) Within that, we encode notes and rests in the disk that corresponds to their duration. 
    Also, we fill the arc at the note's or rest's offset with (b) the color corresponding to their duration. These illustrations are only suitable for simple time signatures.
    }
\end{figure*}

\section{Rhythmic Fingerprint Design}
\label{sec:design}

Our proposed \textit{rhythmic fingerprint} exploits the clock metaphor~\cite{Guo2017}, a psychological anchor~\cite{Vessey1991} in human cognition, and the rhythm tree~\cite{Longuet-Higgins1979}, a music theory concept, to reflect rhythmic aspects. 

\subhead{Characteristics of Rhythm} -- 
\label{sec:rhythm-characteristics}
Occasionally, the term of \textit{rhythm} synonymously represents the phenomena of meter, accent, and timing~\cite{London1908}.
In this work, we define rhythm as referring to two primary dimensions: \textit{onset} and \textit{duration}~~\cite{Krumhansl2000}. 
We consider notes as the presence of sound and its absence as rests. 
To quantify onset, we operate within the context of a single measure and calculate according to a predefined tatum \cite{Mueller2015TempoBeatTracking} of 32\textsuperscript{nd} notes. 
Typically, the duration of a note scales with a predefined tactus~\cite{Mueller2015TempoBeatTracking} of quarter notes depending on the time signature.

There are diverse types of time signatures such as simple, compound, complex, and additive meter~\cite{rashid2018music}. To address all specific attributes of each time signature type exceeds the scope of this paper. The following detailed description addresses the \textit{simple time signature}, but we later explain in \autoref{sec:discussion} how we could extend our approach to deal with other time signatures like \textit{compound meter}.

A fundamental quality of duration is hierarchy \cite{Longuet-Higgins1979}. 
This characteristic emerges in the rhythm tree as depicted in \autoref{fig:rhythm-tree-original}. 
The rhythm tree is similar to the \textit{phrase structure tree} that Chomsky introduced for the syntax analysis in linguistics~\cite{Chomsky1985}, due to the structural similarities between language and music~\cite{Bershadskaya2015}. 

Another characteristic of the rhythm tree is the geometric progression of note durations~\cite{Stone1963}. 
The \textit{duration} of notes adheres to
    \begin{equation}
        d(i) = r^i
        \label{eq:duration-progression}
    \end{equation}
where 
$i\in(-\infty, 2]\,\cap\,\mathbb{Z}$ 
and $r = 2$. 
Here, $i$ represents the \textit{index} of a layer in the rhythm tree (see~\autoref{fig:creating-rhythmic-fingerprint-rhythm-tree}) and $r$ reflects the common \textit{ratio} between successive layers in the rhythm tree (i.e., the duration length bisects at every step from top to bottom). 
For example, the longest note (i.e., a whole note) has a duration of $d(2) = 4$, 
whereas the shortest note (i.e., a 32\textsuperscript{nd} note) has a duration of $d(-3) = 0.125$. 
As discussed, rhythms that deviate from this principle, such as dotted notes and triplets, are not covered by the geometric progression.

\newpage
\noindent The rhythm tree represents the complexity of rhythm in a measure as the distribution of rhythmic contents in the tree's hierarchy. This measure of complexity is similar to the Kolmogorov complexity~\cite{Hashemi2008}. 
\begin{wrapfigure}[4]{r}{0.2\linewidth}
 \vspace{-15pt}
    \begin{center}
    \hspace{-0.7cm}
    \includegraphics[width=1.3\linewidth]{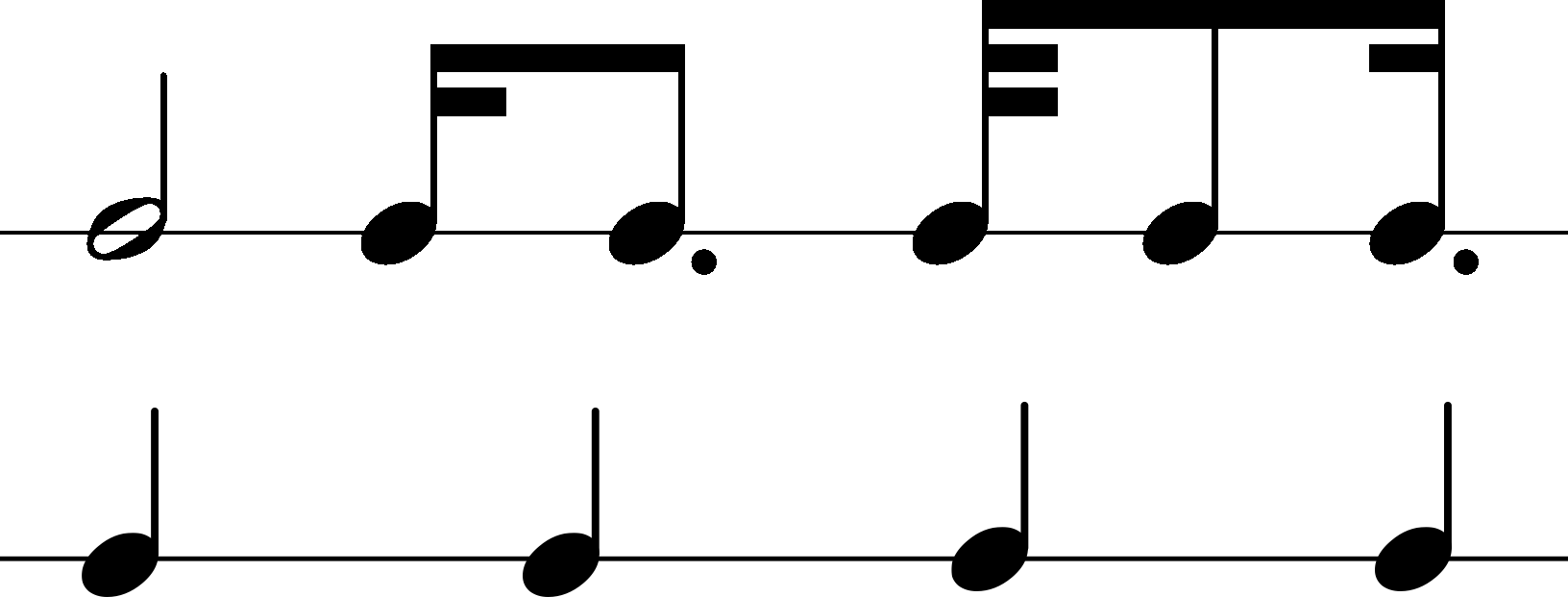}
    \end{center}
\end{wrapfigure}
For example, the first rhythm (1) is more complex than the second rhythm (2), which is simply repeating a quarter note four times. 
Such complexity of rhythm affects the understanding and performance of musical pieces~\cite{Thul2008}.
Related to this complexity is the measurement of evenness, which refers to onsets equally distributed over time~\cite{Toussaint2010}. 
In this example, rhythm (2) has a higher evenness than rhythm (1) due to identical inter-onset intervals. 
The \textit{inter-onset interval} measures the temporal distance between two subsequent onsets. 
Considering rhythmic evenness helps to classify rhythmic patterns~\cite{Toussaint2010}. 
Maximal evenness depends on whether all inter-onset intervals are equal, as it is the case for rhythm (2). 
In this case, the rhythm is even mirror-symmetrical. 
The notion of symmetry is another relevant factor in the analysis of the structure and rhythm \cite{Kempf1996, Toussaint2012}.
Thus, considering symmetry and evenness for visual designs supporting rhythmic analysis is essential.

\subhead{Design Rationale} --
Reading a radial clock is a common task that is anchored in human intuition~\cite{Vessey1991,Guo2017}. 
Fuchs et al. show the advantage of this representation for identification and comparison tasks compared to other representations~\cite{Fuchs2013}. 
Therefore, we use a radial template~(see~\autoref{fig:creating-rhythmic-fingerprint-empty}) for the proposed rhythmic fingerprint design that is based on the clock metaphor~\cite{Guo2017} to reflect the temporal and repetitive aspects of rhythm. 
In the circular design, we layout the content of a measure by starting at the top center continuing in a clockwise direction. 
Then we map the structure of the rhythm tree to this layout~(see~\autoref{fig:creating-rhythmic-fingerprint-rhythm-tree}). 
Since a whole note has the longest duration, 
one traversal of the circle represents 
a duration of $d(2) = 4$.

Our proposed design uses a concentric disk arrangement, similar to a fan chart to radially depict the duration layers of the rhythm tree. 
The innermost disk represents the top layer~(i.e., semibreve) of the rhythm tree (see~\autoref{fig:creating-rhythmic-fingerprint-rhythm-tree}). 
The number of layer nodes, depicted as arcs, is given by
    \begin{equation}
        a(i) = \frac{t}{d(i)}
        \label{eq:arc-count}
    \end{equation}
where 
$i\in(-\infty, 2]\,\cap\,\mathbb{Z}$ 
, $t = 4$, and $d$ is the function given in~\autoref{eq:duration-progression}. 
Here, $i$ represents the layer \textit{index} in the rhythm tree (see~\autoref{fig:creating-rhythmic-fingerprint-rhythm-tree}) and $t$ the \textit{tactus}-level of a quarter note. Since $a$ is strictly increasing, there are more outer than inner disk arcs, which is favorable as they have more display space devoted to them. 
We define the tatum to a 32\textsuperscript{nd} note
(i.e., $i\in[-3, 2]\,\cap\,\mathbb{Z}$). 
This general definition allows for a flexible extension to a shorter tatum such as 64\textsuperscript{th} notes if required.

\noindent The starting point of an arc depicts a note's onset. Therefore, the starting point of each disk's first arc is located at the top center within the fingerprint's circle as indicated in~\autoref{fig:creating-rhythmic-fingerprint-measure}. 
For the remaining arcs, we retrieve the \textit{starting points} according to
    \begin{equation}
        s(i) = \bigg\{\frac{c}{x}\,\Big\vert\,x\in[1,\ a(i)]\,\cap\,\mathbb{Z}\bigg\}
        \label{eq:onsets}
    \end{equation}
where 
$i\in(-\infty, 2]\,\cap\,\mathbb{Z}$ 
, $c = 360^{\circ}$, and $a$ is the function given in \autoref{eq:arc-count}. 
Here, $i$ represents the \textit{index} of a disk and $c$ the \textit{circumference} of the circle. 
The space between two starting points in a disk visually reflects the duration of the arc beginning at the former starting point.

\begin{figure}[t]
\includegraphics[width = \linewidth]{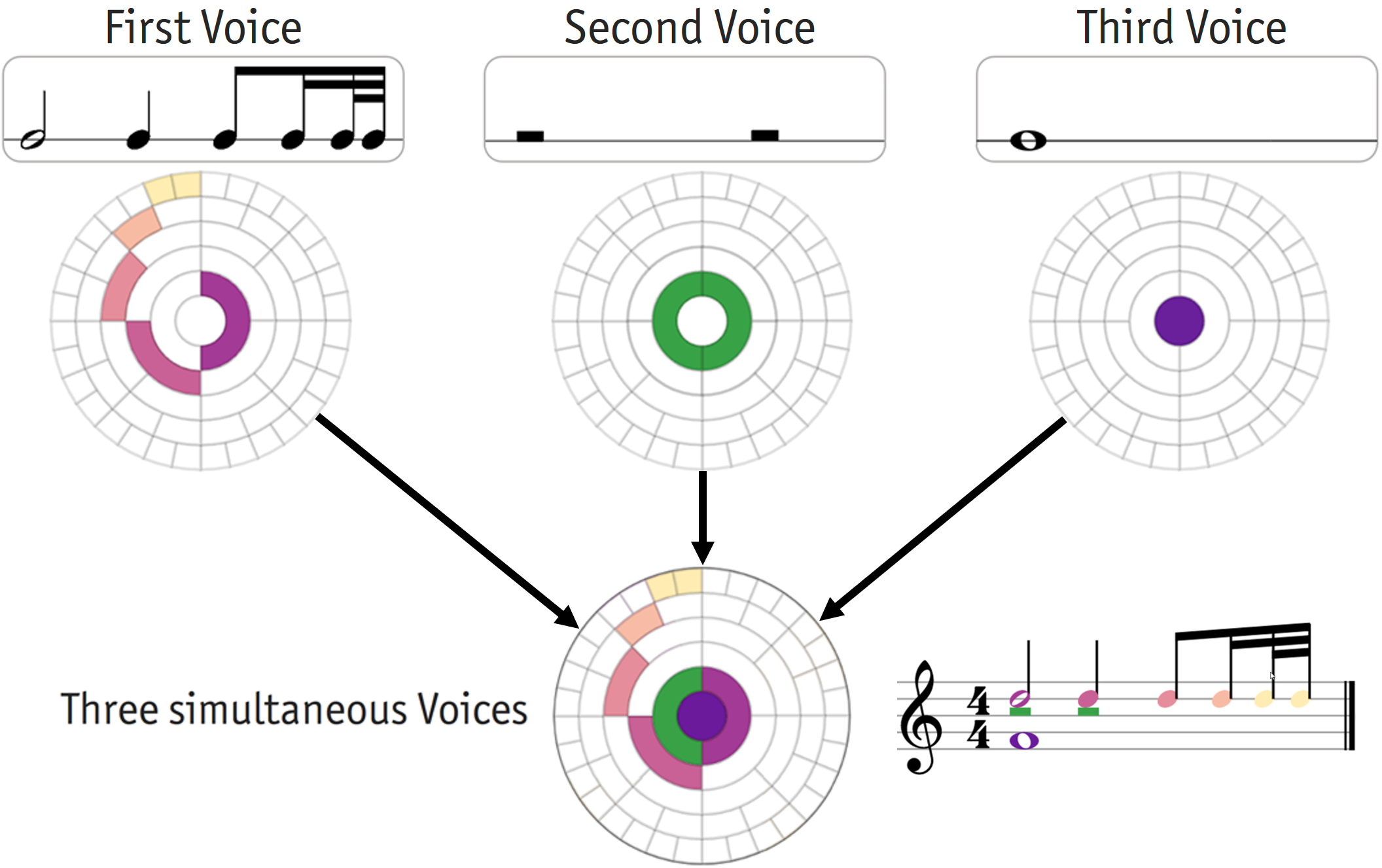}
	\caption{ 
    	The rhythmic fingerprint encodes simultaneous voices. 
    	The \textit{first} (notes) and \textit{second} voice (rests) do overlap in the \textit{first half} segment. 
    	Our design favors notes over rests (here, the first half note). 
    	The third voice has no overlap with the other voices.
	}
	\label{Fig:voiceseparation}
\end{figure}

To convey the rhythmic contents of a measure, we color the arcs using the color scales illustrated in~\autoref{fig:creating-rhythmic-fingerprint-rhythm-tree}. 
The outer disks use a lighter color to avoid a visual bias towards their arcs since shorter notes tend to appear more often than longer notes.
The exemplary measure in~\autoref{Fig:voiceseparation} is a constructed composition of notes and rests to illustrate how three voices 
are encoded by the proposed rhythmic fingerprint.
The third voice plays a whole note that fills the innermost disk, while the second voice comprises two half rests filling the second disk. 
The first voice contains multiple notes of decreasing duration encoded through arcs in different disks. 
This example illustrates the conflict between a note and a rest of the same duration at the same offset (i.e., a half note and a half rest at offset 0). 
As both would address the same arc, we prefer to show the presence of sound rather than its absence. 
Similarly, we aggregate multiple notes or rests of the same duration at the same offset since their multiplicity does not change the perceived rhythm~\cite{Leve2011}.

\noindent Apart from the durations and onsets covered by \autoref{eq:duration-progression} and~\autoref{eq:onsets}, music contains further rhythmic phenomena, such as dotted notes whose duration is given by the extension of~\autoref{eq:duration-progression}:
    \begin{equation}
        d_n(i) = \sum_{k=0}^{n} d(i - k)
            \end{equation}
where 
$i\in(-\infty, 2]\,\cap\,\mathbb{Z}$.
Here, $n$ reflects the \textit{number} of dots that extend a note, $i$ represents the layer \textit{index} in the rhythm tree, and $d$ is the function given in~\autoref{eq:duration-progression}.
There are two additional mechanisms the rhythmic fingerprint utilizes to accommodate those appearances. 
We visualize a dotted note whose base duration belongs to disk $i$ in disk $i + 1$ to emphasize the note's extended duration. 
Thus, a dotted quarter will reside within the half notes disk, for example.
\newpage
\subhead{Augmenting Sheet Music with Rhythmic Fingerprints} -- 
{{MusicXML~\cite{Good2001}}} is a standard file format that is used 
to share digital sheet music and also provides layout information. 
MusicXML has been widely adopted due to the rise of services like MuseScore\footnote{https://musescore.com/}, IMSLP\footnote{https://imslp.org/}, and their joint initiative, OpenScore\footnote{https://openscore.cc/} which distributes compositions in the MusicXML format. 
We leverage this advantage to extract rhythm features, including onset and duration for all notes in every voice, from music sheets using \texttt{music21}~\cite{Cuthbert2010}. 
At the same time, we render the MusicXML with OpenSheetMusicDisplay\footnote{https://opensheetmusicdisplay.org/} (OSMD), an open-source library as a Scalable Vector Graphics. 
By extending OSMD with D3.js~\cite{Bostock2011}, we can place the rhythmic fingerprints on top of the score. 
To ensure flawless augmentation, we enlarge the space between the musical systems to position the rhythmic fingerprints without overlapping with the CMN.
\section{Use Cases}
\label{sec:usecases}

We present three use cases supported by the rhythmic fingerprint introduced in \autoref{sec:design}. 
These use cases illustrate how the rhythmic fingerprint design facilitates music analysis and interpretation tasks.

\subhead{Interpretation of Rhythm} --
Rhythm exhibits a multitude of characteristics, as we outline in \autoref{sec:related-work}. 
To judge a composition's complexity, analysts need to examine each measure to extract the encoded rhythm. 
The rhythmic fingerprints support this task as they enable judging the complexity based on the double encoding of rhythm through color and position.
Higher color diversity indicates a higher complexity, while the opposite holds for a narrow color spectrum. 
Along with the complexity of rhythm comes its evenness. 
This feature depends on onset and duration, which are difficult to measure, especially in polyrhythms. 
Nonetheless, evenness is vital to compare rhythmic patterns and classify music based on it. 
The rhythmic fingerprints support this task as they reflect evenness in their visual patterns. 
For performance preparations, a reader could be interested in finding out if and where a composition is lively or slow.
The color distribution of the rhythmic fingerprints aids this task, especially for longer sheet music, where a majority of darker colors hint at slow music while lighter colors at the outer layers indicate the opposite.

\subhead{Music Structure Analysis} -- 
In \autoref{sec:related-work}, we discussed how rhythm affects the structure of music.
The repetition of and the change in rhythmic patterns constitute section boundaries.
To detect rhythmic changes and repetition, an analyst would need to closely analyze single measures by examining all notes to compare them with the content of other measures.
Typically, this manual analysis process is tedious and time-consuming. 
The annotated rhythmic fingerprints can be of help through the visual patterns.
Instead of delving into the CMN, the rhythmic fingerprints guide the user to recurrent rhythmic patterns.
By keeping the connection to the CMN, analysts can verify hypotheses generated based on the fingerprints.

\subhead{Comparative Analysis of Compositions} --
The rhythmic fingerprints facilitate the comparative analysis of multiple compositions through the combination of rhythm characteristics and music structure.
Instead of individually comparing measures and their notes across compositions, the rhythmic fingerprints aggregate multiple voices into visual patterns.
The combination of their consistent structure and color encoding establishes a visual appearance of measures as single units that are more accessible to an analyst.
They can visually match the exhibited patterns to identify rhythmic similarities, the characteristics of the present rhythms, and music structure. 
The rhythmic fingerprints enable pattern matching due to their preattentive nature using color and consistent positioning.
Comparing two scores with rhythmic fingerprints, a music analyst can distinguish a fast introduction from a slow opening, for example. The former features short-lasting notes which appear in the outer disks of the rhythmic fingerprints, creating bright ring-like appearances. The slow-paced opening consists of longer notes that the rhythmic fingerprints display at their center, creating a dark and compact representation. The analyst can conclude the difference between the two pieces of music at a glance, easing the comparative analysis.

\begin{figure*}[ht]
    \centering
    \subfigure[][
        We asked \noviceB~to highlight recurring patterns in \MSA~\textit{without} the rhythmic fingerprints. 
        Besides identifying pattern $D$ by three sub-patterns, he could only identify small parts (e.g., in $A$) or none (e.g., $C$) of the other patterns.
            ]{\includegraphics[width = 0.4725\linewidth]{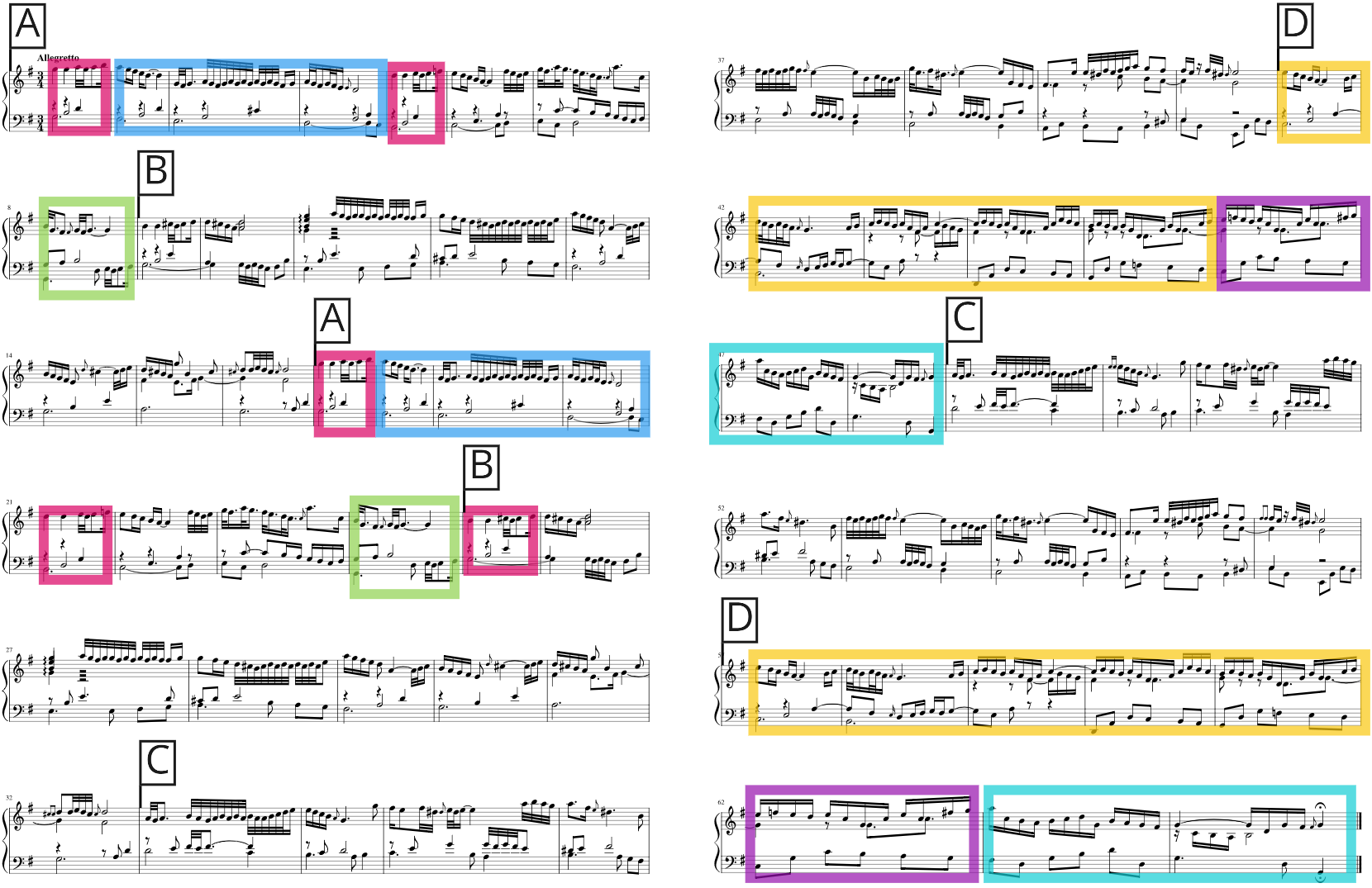}\label{fig:nov2-results-cmn}}
    \hfill
    \subfigure[][
        We asked \noviceB~to highlight recurring patterns in \MSB~\textit{with} the rhythmic fingerprints. 
        He came up with the pattern $C$, treated $AB$ as a single repeating pattern, and found multiple nested patterns throughout larger patterns.
    ]{\includegraphics[width = 0.4725\linewidth]{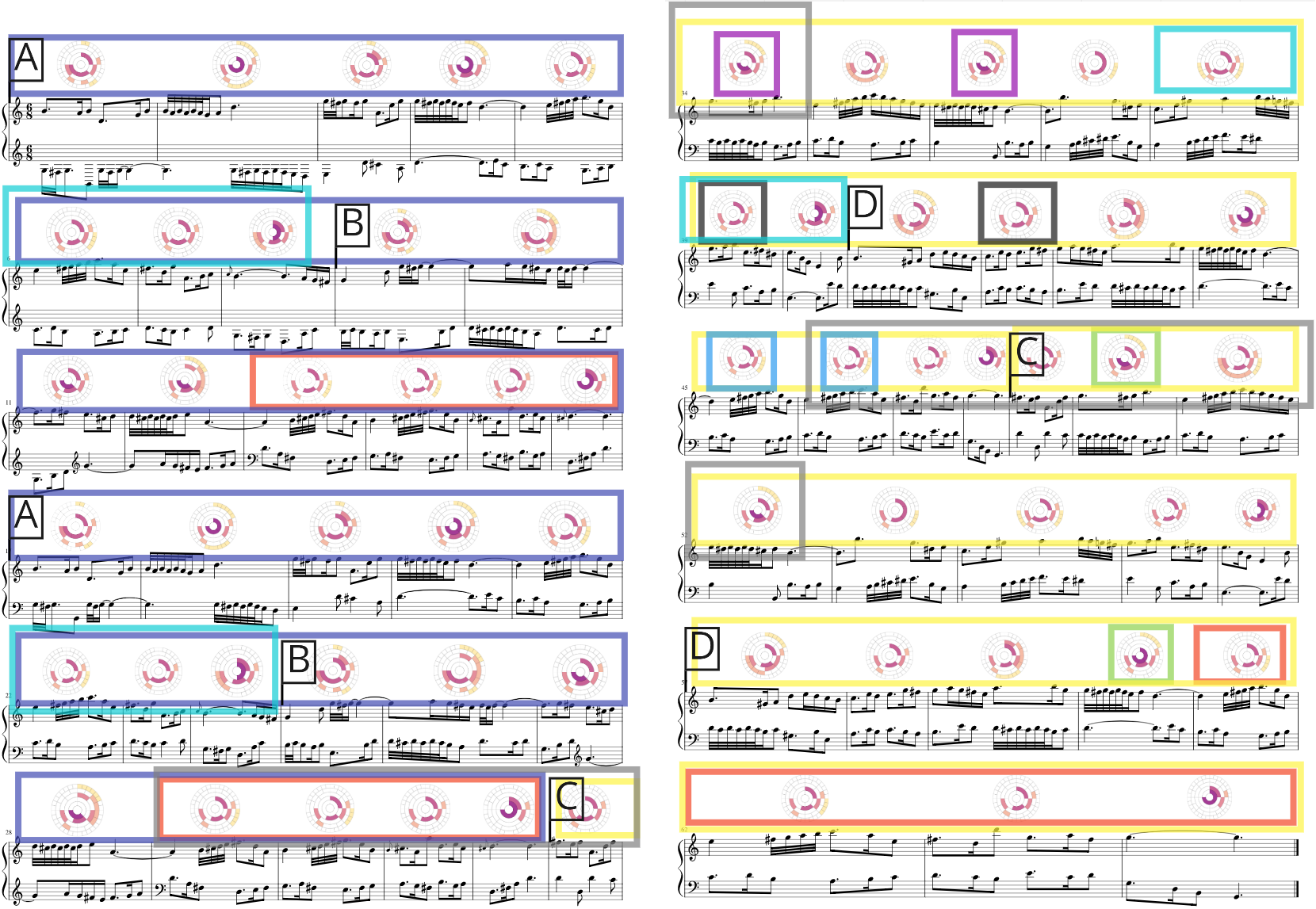}\label{fig:nov2-results-fingerprints}}
    \caption{
        During the user study, we instructed the participants to find and mark as many recurring patterns as possible by colored rectangle annotations within the presented music sheet using the CMN (a) without and (b) with the rhythmic fingerprints augmentation.
    }
    \label{fig:nov2-results}
\end{figure*}

\section{Evaluation}

To evaluate the rhythmic fingerprint introduced in \autoref{sec:design}, we conducted a qualitative user study similar to our previous work~\cite{Miller2019}. 
We tasked each participant to identify patterns in two music sheets, one with the rhythmic fingerprints and the second one without. 
This approach enables us to assess the advantages of the rhythmic fingerprint augmentation and elicit qualitative feedback from the users.
To highlight differences between novices and experts, we divided the participants into two separate groups based on their expertise level regarding music theory knowledge and rhythm analysis.

\subsection{Study Methodology and Design}
Our qualitative assessment of the rhythmic fingerprints comprises three steps: 
(1) The introduction of the design, 
(2) the analysis with and without the fingerprints, and 
(3) gathering feedback during the analysis. 
Hereinafter, we describe the used datasets, the participant characteristics, and the study procedure.

\subhead{Data Sets and Ground Truth} --
During the course of this study, the participants analyzed two music sheets, \textit{Aria} (\MSA) \cite{GoldbergVariationsMusescore} and \textit{Variation VII} (\MSB) \cite{GoldbergVariationsMusescore} from the \textit{Goldberg Variations} by Johann Sebastian Bach~(see \hyperref[sec:appendix]{Appendix}).
To ease the assessment of the analysis task, we exploit available ground truth \cite{Breig1975}.
Both music sheets \MSA~and \MSB~are equally long, consisting of 64 measures, and have the time signatures \setmeterb{3}{4} and \setmeterb{6}{8}, respectively. 
We chose these compositions due to their comparable complexity and musical form. 
The 64 measures are divided into two parts, namely $P_1$ and $P_2$. 
Each part covers 32 measures and consists of a repeated pair of sections: $AB$ for $P_1$ and $CD$ for $P_2$ with a resulting form $ABABCDCD$ as indicated by the lettered labels in~\autoref{fig:nov2-results}. 
This overall structure is reflected by the rhythmic patterns, allowing us to compare the results between the unmodified CMN and the augmented music sheet.

\subhead{Participants} --
We conducted our study with eight participants that we divide into two groups of different expertise levels. 
The first group consists of four \textit{novices} ($N_1$ - $N_4$) who have little experience in music theory (mean score of 2) and never performed rhythm analysis before (mean score of 1). 
The second group includes four \textit{experts} ($E_1$ - $E_4$) who have more advanced experience in music theory (mean score of 2.5) and are more proficient in rhythm analysis (mean score of 3.25). 
All participants in both groups have a university degree and are aged $29.5\pm 4.5$ on average.

\subhead{Procedure} -- 
The evaluation comprised three phases. 
First, the participants provided demographic information and their level of expertise
using a Likert scale from beginner~(1) to expert~(5).

During the second phase, we familiarized the participants with the rhythmic fingerprint design. 
Then, we verified their understanding of the explanations by presenting examples each participant had to solve correctly. 
We continued explaining the rhythmic fingerprint based on an exemplary music sheet to introduce polyphonic rhythms they would encounter during the analysis.

In the last study step, the participants analyzed two music sheets: one with and another without the rhythmic fingerprints. 
We also randomized the condition order
such that each condition applies to one novice and one expert.
For every condition, we asked the participant to identify recurrent rhythmic patterns that are exact matches throughout all voices, limiting each study pass to 30 minutes. 
Eventually, we elicited feedback regarding the analysis and the rhythmic fingerprints, depending on the condition, during an interview. 
We repeated the same process for the second pass and concluded with a comparison of both analysis sessions in the final interview to learn about strategies for the given comparison task.

\subsection{User Feedback}
We elicited qualitative feedback from each participant and assessed their performance by comparing their results with the ground truth.

\subhead{Pattern Identification Strategies} -- 
During the analysis, the participants followed diverse strategies to identify rhythmic patterns.
Without the rhythmic fingerprints, most of the participants focused on a single stave. 
\noviceB~stated that he ``\textit{focused on the second voice because it was easier as it contained fewer notes}''. 
Meanwhile, \noviceC~started differently and ``\textit{looked at the melody voice, because that's the most distinctive}''. 
Later, both combined their findings with the respective other staff to match rhythmic patterns. 
\expertC took another approach as they argued that ``\textit{16\textsuperscript{th} and 32\textsuperscript{nd} notes are [...] quite memorable or quarters as a starting point, they do not occur [...] often}''. This idea concurs with \expertD~who found that ``\textit{32\textsuperscript{nd}s always stand out extremely}''. 
This strategy focuses on salient features of the visual appearance created by the CMN to form rhythmic patterns.

\noindent Given a music sheet with the rhythmic fingerprint augmentation, the participants, except for \noviceA~and \expertA, favored the fingerprints to spot recurrent rhythmic patterns.
Consequently, many of the users focused on the rhythmic fingerprints while neglecting the CMN. 
Still, \expertD~``\textit{tried to double-check [the results]}'' and referred to the CMN as means of verification. 
To identify recurrent rhythmic patterns with the rhythmic fingerprints, the participants preferred salient aspects of the fingerprints. 
As examples they gave the ``\textit{dark colour [which] is striking}'' (\noviceD), ``\textit{long rhythmic runs}'' (\expertA), the ``\textit{yellow on the outside}'' (\noviceB), and ``\textit{the green colour [which] is [...] prominent}''~(\expertC).

\subhead{Usefulness of the Rhythmic Fingerprints} -- 
When asked how the augmentation supported the analytical tasks, the participants gave various insights. 
All of them, except for \noviceA, appreciated the rhythmic fingerprints as they eased the analysis. 
Not only ``\textit{have [they] greatly accelerated the analysis}'' (\noviceC) but they ``\textit{create visual[ly] clear patterns}'' (\noviceB) such that one ``\textit{could easily find the similarities}'' (\noviceB) and 
``\textit{larger continuous patterns}''~(\noviceC). 
During his first analysis with the CMN alone (see \autoref{fig:nov2-results-cmn}), \noviceB~did not manage to find any ground truth patterns.
By contrast, in his second analysis with the rhythmic fingerprints (see \autoref{fig:nov2-results-fingerprints}), he was able to identify the two parts, $P_1$ and $P_2$, as well as their repeating sections $AB$ and $CD$, respectively. 
In need of the next analytical step, \expertA~fittingly noticed that ``\textit{a fingerprint would be helpful for a new push because it was much easier to look for things that stand out}'' while analyzing the CMN alone. 
\expertC~even goes so far as to say that ``\textit{someone who has never dealt with rhythm before [could] still do an analysis of rhythm here}'' and ``\textit{[wouldn't] need to know notes}''.

\subhead{Challenges} -- 
None of the participants encountered the rhythmic fingerprint before our evaluation. 
The participants faced different challenges during the analysis. 
At start, many participants ``\textit{found it extremely difficult to get used to the fingerprint}''~(\expertB). 
Apart from the steep learning curve, the participants struggled with the current color usage. 
Particularly, \expertC~found that ``\textit{the colors are too similar, especially the red [ones]}'' which \expertB~emphasized, concerned that ``\textit{the 32\textsuperscript{nd}s are difficult to see}'' in contrast to the white background. 
They further elaborated on the consequence being that ``\textit{it is more difficult to search on the outer circle compared to the inner one}'' which led \noviceC~to ``\textit{ look at them almost as closely as the notes}'', hampering the usefulness of the rhythmic fingerprints. 
If not looked at them closely enough, \noviceA~and \expertA~feared ``\textit{a likelihood of confusion because the circles look very similar}'' (\noviceA). 
\noviceC~and \expertD~experienced a similar irritation as they ``\textit{lost the context [when only focused] on the glyphs}'' (\noviceC) and had a hard time ``\textit{to remember the fingerprints and to find them somewhere else}'' (\expertD). 

\section{Discussion}
\label{sec:discussion}
Based on the study results we assess the pattern identification performance of the participants.
With the fingerprints, each participant had a similar or better performance than with the CMN only.
Since we provided an elaborated introduction on the rhythmic fingerprints before performing the actual analysis task, both novices and experts understood and thus profited from the rhythm visualization.
This familiarity could be a reason for the trust of the users towards the rhythmic fingerprints. 
Although \noviceA~and \expertA~primarily relied on the CMN, the remaining participants mainly worked with the visualization.
Although some users reported a steep learning curve for the rhythmic fingerprints, \noviceA~argued that the use of color may inspire musicians to analyze the rhythm. 
\expertA~added that the rhythmic fingerprints could give a new impulse during the rhythm analysis.

\subsection{Lessons Learned} 
The main discussion points gathered during the rhythm analysis process and the user's feedback are: 
(1) the fingerprint inherently encodes rhythmic complexity, 
(2) the fingerprint does not separate between different voices, and 
(3) the fingerprint encodes a limited number of time signatures.

\subhead{Rhythm Complexity} -- 
The rhythm tree and the rhythmic fingerprints, express complexity through the diversity of onsets and durations. 
\expertB~accurately noticed that a greater variety corresponds to higher complexity. 
Therefore, the amount of present notes influences readability. 
In the final interview, \expertC~reported that due to the limited space of the inner disks and the adjacent placement, it was challenging to distinguish the colored arcs. 
Analogously, \expertA~stated that due to the smaller size, the outer disks were more challenging to compare than the inner disks.
While the CMN enables music readers to view multiple voices and their rhythmic progression separately, the rhythmic fingerprint aggregates the commonalities and reveals the rhythm perceived by listeners. 
Consequently, the fingerprints do not replace the well-established CMN, which can be used by analysts to understand the details of multiple voices.
We also learned that it is currently difficult to see the arcs at the outer disk due to the small lightness difference to the background. 
To address this issue or other visual impairment issues, it is no effort to change the color scale according to the reader's needs.

\subhead{Voice Separation} --
In a multi-voice setting, the current design does not support readers distinguishing separate voices as in the CMN.
To achieve this, readers need to examine the CMN to understand voice separation.
Due to our decision to emphasize the presence of notes over rests in cases of overlap with the same offset and duration, the rhythmic fingerprint aggregates simultaneous notes and rests. 
As outlined in \autoref{sec:design}, we favor the presence of sound over its absence.
Consequently, the rhythmic fingerprint is not bidirectional. 
It visually simplifies parallel rhythmic content to extract perceived rhythm, which is not altered by the multiplicity of notes with identical onset and duration.
Due to this simplification, the rhythmic fingerprint design does not support multiple voice analysis. 
Tailoring the design to support such tasks would increase the design complexity since more information would be represented.

\subhead{Time Signature} -- 
One traversal of the rhythmic fingerprint represents a whole note's duration. 
Thus, even longer durations cannot be displayed correctly. 
Simple time signatures, such as common, \setmeterb{3}{4}, or cut time, can be analyzed without any issues. 
\expertB~noticed that the design of the rhythm visualization allowed him to determine a composition's time merely by looking at the appearance of the rhythmic fingerprint. 
Meanwhile, none of the other participants faced issues during their analysis due to our design choice.
Contrary, the visualization does not properly accommodate all possible time signatures, such as \setmeterb{5}{4}. Since such time signatures would require more than one traversal, only two instances of the rhythmic fingerprint could adequately display the rhythmic content of such a measure.

\subsection{Future Work}
While the study showed that the rhythmic fingerprints support the analysis of rhythmic patterns even when used by novice readers, there are still restrictions and limitations of the proposed approach. 
First, there are numerous time signatures of varying complexity~\cite{rashid2018music}. 
The most common time signatures are simple (e.g., \setmeterb{3}{4}) and compound (e.g., \setmeterb{6}{8}) meter. The current fingerprint skeleton (see~\autoref{fig:creating-rhythmic-fingerprint-empty}) only addresses simple meters.
\begin{wrapfigure}[11]{r}{0.2\linewidth}
 \vspace{-18pt}
    \begin{center}
    \hspace{-0.7cm}
        \includegraphics[width=1.2\linewidth]{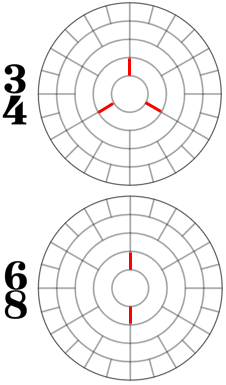}
    \end{center}
\end{wrapfigure}
While it is theoretically possible to visualize compound meters with the current design, it does not properly accommodate for the tertiary grouping, e.g., three eighth notes that are encompassed by a dotted quarter note.
The two skeletons to the right show how \setmeterb{3}{4} and \setmeterb{6}{8} differ (i.e., red lines in the 2\textsuperscript{nd} inner disk) regarding their skeletons.  The main difference between the duple and the triple meter is the perception of emphasized beats in a measure~\cite{Longuet-Higgins1979}, which is especially relevant for performers who need to know which notes to emphasize. Typically, in \setmeterb{6}{8} meter, dotted quarter notes are indicating the regular pulse of the composition.

There are different directions in which to extend the proposed design further.
We plan to research how the design can support other time signatures~\cite{rashid2018music} with the additional use of interaction techniques. 
Consequently, we want to tailor the visualization to the individual needs of music analysts to overcome the current restrictions of static rhythmic fingerprints.
For instance, \textit{Detail Outside} and \textit{Detail Inside} are interaction techniques designed to increase the readability in dense fan charts~\cite{Stasko2000}. 
We want to provide a visual music analysis interface that integrates such interaction opportunities with digital sheet music to facilitate the access based on MusicXML~\cite{Good2001}.
Moreover, we plan to extend the temporal range for the fingerprints from single measures to larger sections of a musical composition to provide overviews about the rhythmic content. 
With this, we take the view that music analysis can further benefit from visualization research for close and distant reading, an important concept in digital humanities~\cite{Janicke2015}. 
To investigate the potential of other designs, we plan to organize design workshops and competitions to elicit new ideas for further musical visualizations suitable for music analysis tasks.

\section{Conclusion}
We proposed a visualization for rhythm by exploiting a concept of music theory, the rhythm tree, to extend traditional music notation.
The augmentation of sheet music with the {rhythmic fingerprints} combines typical close reading with distant reading of sheet music.
To evaluate our approach, we conducted a qualitative user study with four novices and four experts.
The user performance assessment indicates that our visualization improves the identification of rhythmic patterns. During the analysis, users were able to determine the music structure and characteristics of rhythm with the help of the rhythmic fingerprints.
Through the participants' feedback, we also identified difficulties that analysts face while working with the proposed rhythmic fingerprints. 
These include the readability of complex rhythms, the separation of voices, and the display of irrational or complex time signatures.
We plan to compare different layout strategies and color scales in future work to address these challenges.
We aim to combine the rhythmic with the harmonic fingerprints from our previous work~\cite{Miller2019}, enabling the joint analysis of harmony and rhythm based on MusicXML.

\bibliographystyle{abbrv-doi}
\newpage
\bibliography{references}

\clearpage
\section*{Appendix}
\label{sec:appendix}
\hspace*{0.05\textwidth}\begin{minipage}{0.9\textwidth}
    \centering
        \includegraphics[page=1,width=0.475\textwidth]{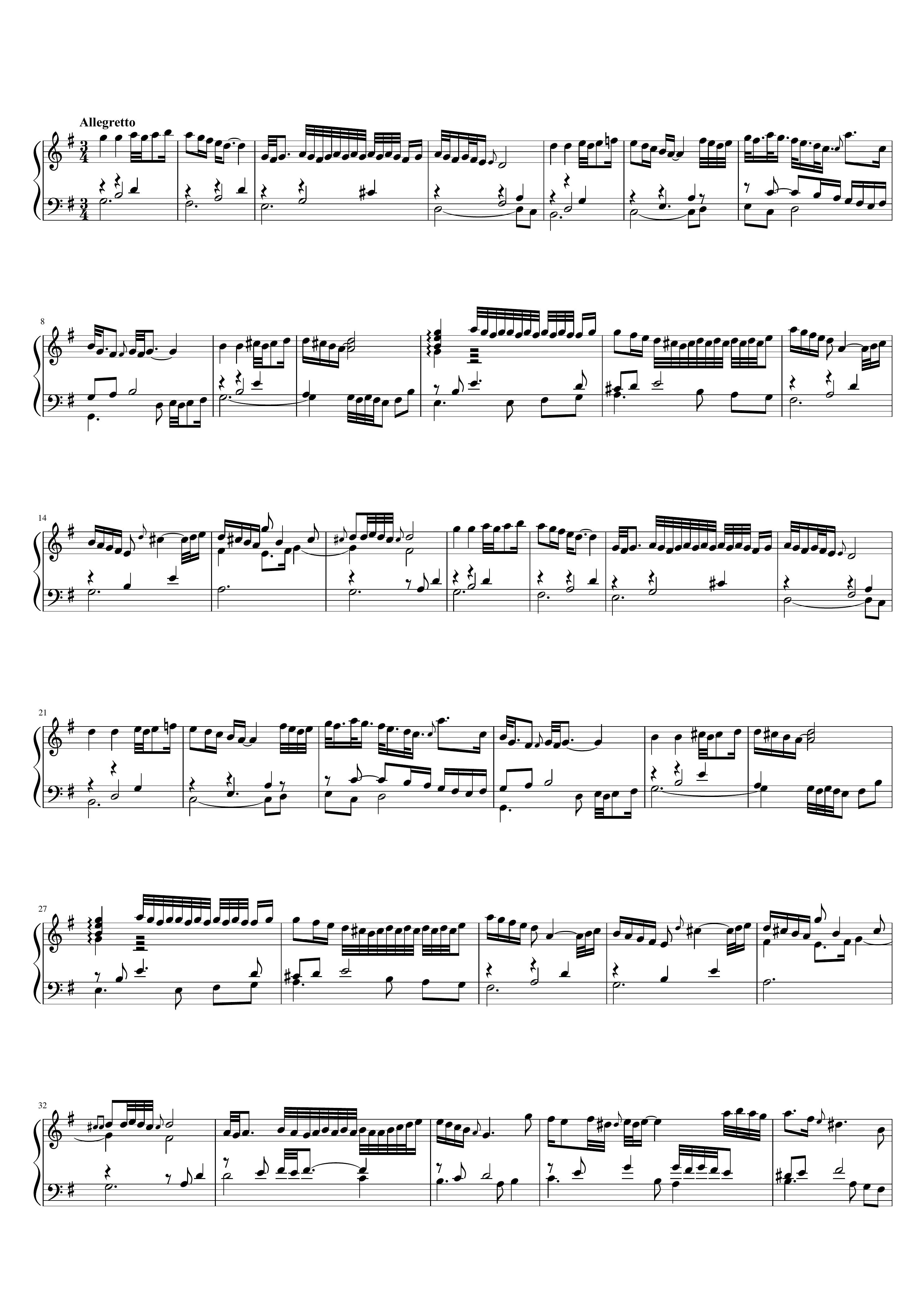}
        \hfill
        \includegraphics[page=2,width=0.475\textwidth]{img/supplemental/Johann_Sebastian_Bach_-_Goldberg_Variations_-_Aria_MS1_without_Fingerprints.pdf}
        
        Figure A.1: The dataset ``Aria'' from the Goldberg Variations by Johann Sebastian Bach without the fingerprints~(\MSA). 
        
        \includegraphics[page=1,width=0.475\textwidth]{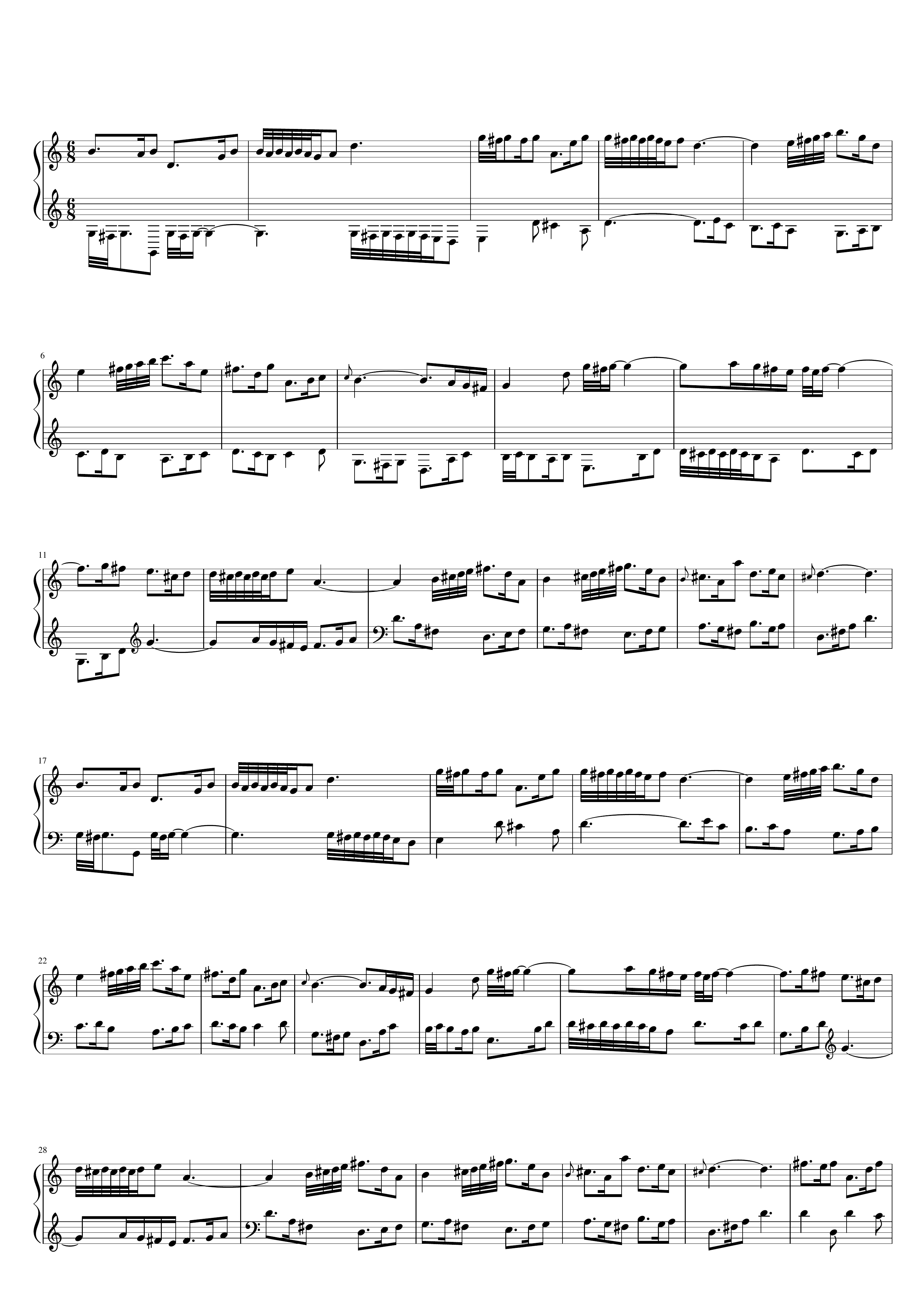}
        \hfill
        \includegraphics[page=2,width=0.475\textwidth]{img/supplemental/Johann_Sebastian_Bach_-_Goldberg_Variations_-_Variation_VII_MS2_without_Fingerprints.pdf}
        
        Figure A.2: The dataset ``Variation VII'' from the Goldberg Variations by Johann Sebastian Bach without the fingerprints~(\MSB). 
\end{minipage} 

\newpage
\ 
\newpage

\hspace*{0.05\textwidth}\begin{minipage}{0.9\textwidth}
        \centering
        \includegraphics[page=1,width=0.475\textwidth]{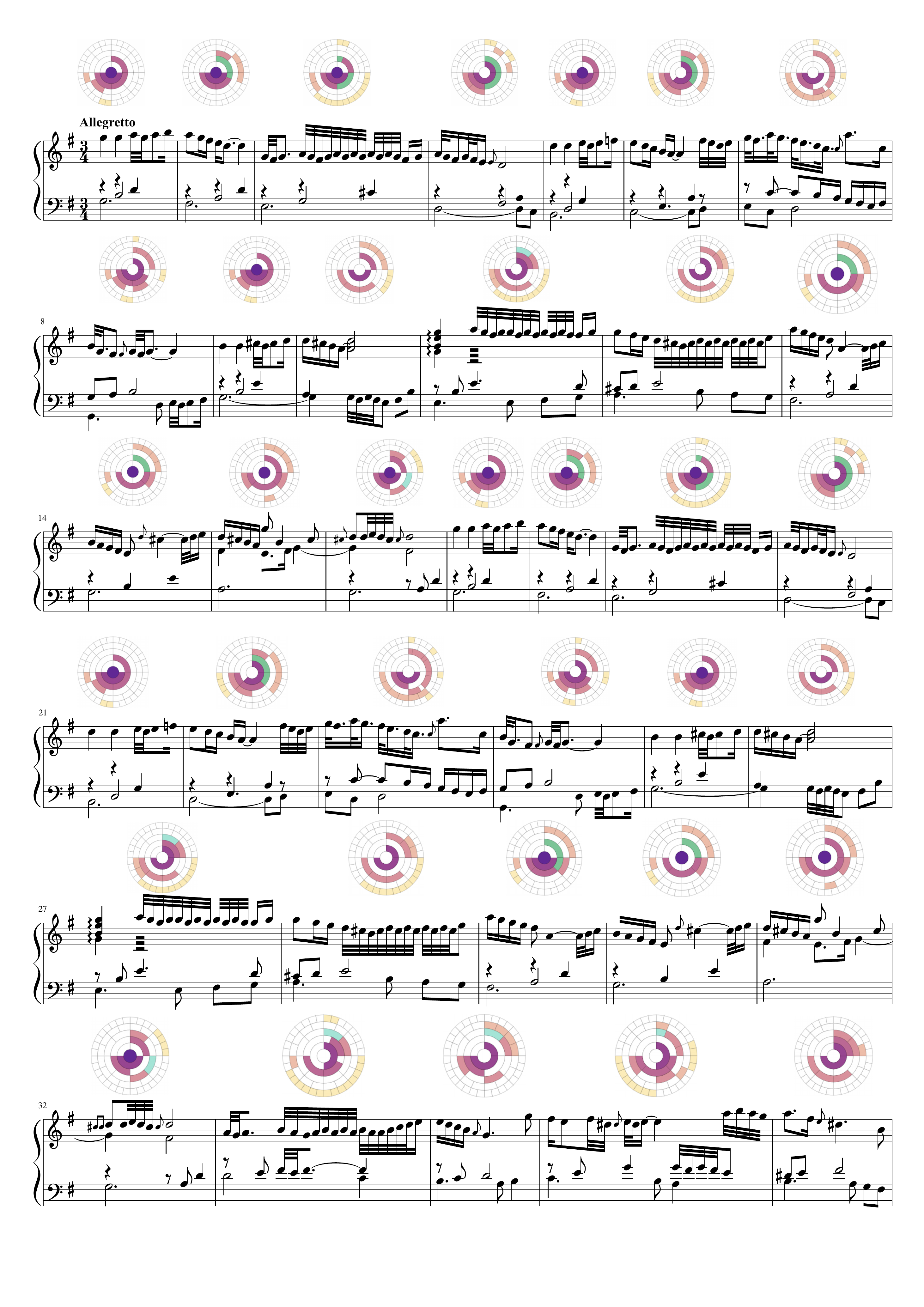}
        \hfill
        \includegraphics[page=2,width=0.475\textwidth]{img/supplemental/Goldberg_Variations_-_Johann_Sebastian_Bach_-_Aria_MS1_with_Fingerprints_optimiert.pdf}
        
        Figure A.3: The dataset ``Aria'' from the Goldberg Variations by Johann Sebastian Bach with the fingerprints~(\MSA). 
        
        \includegraphics[page=1,width=0.475\textwidth]{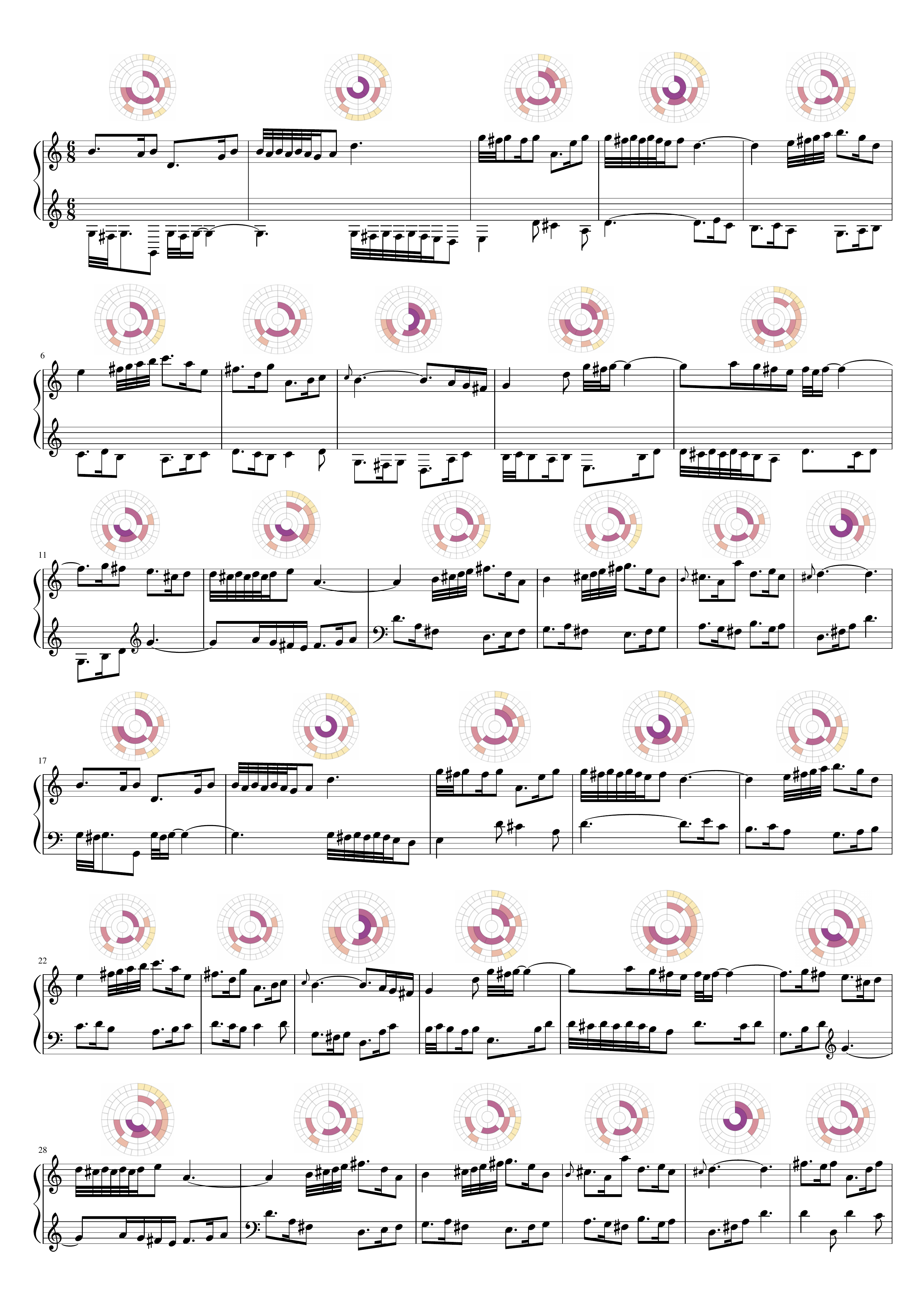}
        \hfill
        \includegraphics[page=2,width=0.475\textwidth]{img/supplemental/Goldberg_Variations_-_Johann_Sebastian_Bach_-_Variation_VII_MS2_with_Fingerprints_optimiert.pdf}
        
        Figure A.4: The dataset ``Variation VII'' from the Goldberg Variations by Johann Sebastian Bach with the fingerprints~(\MSB). 
\end{minipage} 

\newpage
\ 
\newpage

\hspace*{0.075\textwidth}\begin{minipage}{0.85\textwidth}
            \includegraphics[width = \textwidth]{img/new/Ground_Truth_+_N2_Results/Ground_Truth_MS1_+_N2_Results.png}
        
        \vspace{0.5cm}
        Figure A.5: The dataset \MSA~without the fingerprint augmentation annotated by \noviceB. The labels illustrate the underlying ground truth patterns. 
        More examples annotated by the other participants can be found here: \href{https://osf.io/jx8dy/?view_only=fa9345e7a9d8433fa834f6ec97f5359c}{{https://osf.io/jx8dy/}}.
        \vspace*{0.5cm}
        
        \includegraphics[width = \textwidth]{img/new/Ground_Truth_+_N2_Results/Ground_Truth_MS2_+_N2_Results.png}
        
        \vspace{0.5cm}
        Figure A.6: The dataset \MSB~with the fingerprint augmentation annotated by \noviceB. 
        The labels illustrate the underlying ground truth patterns. 
        More examples annotated by the other participants can be found here: \href{https://osf.io/jx8dy/?view_only=fa9345e7a9d8433fa834f6ec97f5359c}{{https://osf.io/jx8dy/}}.
\end{minipage}

\end{document}